%% file: main.tex
\pdfoutput=1
\documentclass[12pt,a4paper]{article}
\usepackage{ifthen} % for conditional statements
\newboolean{pdflatex}
\setboolean{pdflatex}{true} % False for eps figures 

\newboolean{articletitles}
\setboolean{articletitles}{true} % False removes titles in references

\newboolean{uprightparticles}
\setboolean{uprightparticles}{false} %True for upright particle symbols

\newboolean{inbibliography}
\setboolean{inbibliography}{false} %True once you enter the bibliography

\input{preamble}
\usepackage{longtable} % only for template; not usually to be used in PAPERs
\usepackage{graphicx}
\usepackage{subfig}
\usepackage{booktabs}

\begin{document}

%%%%%%%%%%%%%%%%%%%%%%%%%
%%%%% Title     %%%%%%%%%
%%%%%%%%%%%%%%%%%%%%%%%%%
\renewcommand{\thefootnote}{\fnsymbol{footnote}}
\setcounter{footnote}{1}

% %%%%%%% CHOOSE TITLE PAGE--------
%\onecolumn
\input{title-LHCb-PAPER}

%\twocolumn
% %%%%%%%%%%%%% ---------

\renewcommand{\thefootnote}{\arabic{footnote}}
\setcounter{footnote}{0}

%%%%%%%%%%%%%%%%%%%%%%%%%%%%%%%%
%%%%%  Table of Content   %%%%%%
%%%%%%%%%%%%%%%%%%%%%%%%%%%%%%%%
%%%% Uncomment next 2 lines if desired
%\tableofcontents
%\cleardoublepage

%%%%%%%%%%%%%%%%%%%%%%%%%
%%%%% Main text %%%%%%%%%
%%%%%%%%%%%%%%%%%%%%%%%%%

\pagestyle{plain} % restore page numbers for the main text
\setcounter{page}{1}
\pagenumbering{arabic}

%% Uncomment during review phase. 
%% Comment before a final submission.
%\linenumbers

% You can include short sections directly in the main tex file.
% However, for larger papers it is desirable to split the text into
% several semiautonomous files, which can be revised independently.
% This is especially useful when developing a document in
% collaboration with several people, since then different parts can be
% edited independently.  This type of file organization is shown here.
% 

\input{introduction}

\input{detectoranddata}

\input{strategy}

\input{backgrounds}

\input{sensitivity}

\input{systematics}

\input{results}

% Do not include this in analysis note and conference reports
\input{acknowledgements}

%\input{appendix}

% This should be taken out in the final paper
%\input{supplementary-app}

\addcontentsline{toc}{section}{References}
\setboolean{inbibliography}{true}
\bibliographystyle{LHCb}
\bibliography{main,LHCb-PAPER,LHCb-CONF,LHCb-DP,LHCb-TDR}

\newpage                                                 

\input{LHCb_Authorship_flat_28-Feb-2017}
 
\newpage

\end{document}

%% file: preamble.tex
\usepackage[top=1in, bottom=1.25in, left=1in, right=1in]{geometry}
\usepackage{amsbsy}
\usepackage{microtype}
\usepackage{lineno}  % for line numbering during review
\usepackage{xspace} % To avoid problems with missing or double spaces after
\usepackage{caption} %these three command get the figure and table captions automatically small

\usepackage{graphicx}  % to include figures (can also use other packages)
\usepackage{color}
\usepackage{colortbl}
\graphicspath{{./figs/}} % Make Latex search fig subdir for figures

\usepackage{amsmath} % Adds a large collection of math symbols
\usepackage{amssymb}
\usepackage{amsfonts}
\usepackage{upgreek} % Adds in support for greek letters in roman typeset

\newcommand*\patchAmsMathEnvironmentForLineno[1]{%
\expandafter\let\csname old#1\expandafter\endcsname\csname #1\endcsname
\expandafter\let\csname oldend#1\expandafter\endcsname\csname
end#1\endcsname
 \renewenvironment{#1}%
   {\linenomath\csname old#1\endcsname}%
   {\csname oldend#1\endcsname\endlinenomath}%
}
\newcommand*\patchBothAmsMathEnvironmentsForLineno[1]{%
  \patchAmsMathEnvironmentForLineno{#1}%
  \patchAmsMathEnvironmentForLineno{#1*}%
}
\AtBeginDocument{%
\patchBothAmsMathEnvironmentsForLineno{equation}%
\patchBothAmsMathEnvironmentsForLineno{align}%
\patchBothAmsMathEnvironmentsForLineno{flalign}%
\patchBothAmsMathEnvironmentsForLineno{alignat}%
\patchBothAmsMathEnvironmentsForLineno{gather}%
\patchBothAmsMathEnvironmentsForLineno{multline}%
\patchBothAmsMathEnvironmentsForLineno{eqnarray}%
}

\usepackage{hyperref}    % Hyperlinks in references
\usepackage[all]{hypcap} % Internal hyperlinks to floats.

\input{lhcb-symbols-def} % Add in the predefined LHCb symbols
\input{our-symbols-def}

\usepackage{cite} % Allows for ranges in citations
\usepackage{mciteplus}

%% file: lhcb-symbols-def.tex
\usepackage{xspace} 
\usepackage{upgreek}

\def\lhcb {\mbox{LHCb}\xspace}

\def\velo   {VELO\xspace}

\def\MagUp {\mbox{\em Mag\kern -0.05em Up}\xspace}

\ifthenelse{\boolean{uprightparticles}}%
{

 \def\Pmu         {\ensuremath{\upmu}\xspace}                 
 \def\Pnu         {\ensuremath{\upnu}\xspace}                 
                  
 \def\Ppi         {\ensuremath{\uppi}\xspace}

 \def\Ppsi        {\ensuremath{\uppsi}\xspace}

 \def\PDelta      {\ensuremath{\Delta}\xspace}                 
 \def\PXi      {\ensuremath{\Xi}\xspace}                 
 \def\PLambda      {\ensuremath{\Lambda}\xspace}                 
 \def\PSigma      {\ensuremath{\Sigma}\xspace}                 
 \def\POmega      {\ensuremath{\Omega}\xspace}                 
 \def\PUpsilon      {\ensuremath{\Upsilon}\xspace}                 
 
 \def\PB      {\ensuremath{\mathrm{B}}\xspace}                 
                  
 \def\PD      {\ensuremath{\mathrm{D}}\xspace}

 \def\PJ      {\ensuremath{\mathrm{J}}\xspace}                 
 \def\PK      {\ensuremath{\mathrm{K}}\xspace}

 \def\Pb      {\ensuremath{\mathrm{b}}\xspace}                 
 \def\Pc      {\ensuremath{\mathrm{c}}\xspace}

 \def\Pi      {\ensuremath{\mathrm{i}}\xspace}

 \def\Pp      {\ensuremath{\mathrm{p}}\xspace}

}
{

 \def\Pmu         {\ensuremath{\mu}\xspace}                 
 \def\Pnu         {\ensuremath{\nu}\xspace}                 
                  
 \def\Ppi         {\ensuremath{\pi}\xspace}

 \def\Ppsi        {\ensuremath{\psi}\xspace}                 
                  
 \mathchardef\PDelta="7101
 \mathchardef\PXi="7104
 \mathchardef\PLambda="7103
 \mathchardef\PSigma="7106
 \mathchardef\POmega="710A
 \mathchardef\PUpsilon="7107
                  
 \def\PB      {\ensuremath{B}\xspace}                 
                  
 \def\PD      {\ensuremath{D}\xspace}

 \def\PJ      {\ensuremath{J}\xspace}                 
 \def\PK      {\ensuremath{K}\xspace}

 \def\Pb      {\ensuremath{b}\xspace}                 
 \def\Pc      {\ensuremath{c}\xspace}

 \def\Pi      {\ensuremath{i}\xspace}

 \def\Pp      {\ensuremath{p}\xspace}

}

\makeatletter
\ifcase \@ptsize \relax% 10pt
  \newcommand{\miniscule}{\@setfontsize\miniscule{4}{5}}% \tiny: 5/6
\or% 11pt
  \newcommand{\miniscule}{\@setfontsize\miniscule{5}{6}}% \tiny: 6/7
\or% 12pt
  \newcommand{\miniscule}{\@setfontsize\miniscule{5}{6}}% \tiny: 6/7
\fi
\makeatother

\DeclareRobustCommand{\optbar}[1]{\shortstack{{\miniscule (\rule[.5ex]{1.25em}{.18mm})}
  \\ [-.7ex] $#1$}}

\def\mup        {{\ensuremath{\Pmu^+}}\xspace}
\def\mun        {{\ensuremath{\Pmu^-}}\xspace} % muon negative (\mum is taken)
\def\mumu       {{\ensuremath{\Pmu^+\Pmu^-}}\xspace}

\def\ellm       {{\ensuremath{\ell^-}}\xspace}
\def\ellp       {{\ensuremath{\ell^+}}\xspace}

\def\neu        {{\ensuremath{\Pnu}}\xspace}
\def\neub       {{\ensuremath{\overline{\Pnu}}}\xspace}
\def\neum       {{\ensuremath{\neu_\mu}}\xspace}
\def\neumb      {{\ensuremath{\neub_\mu}}\xspace}
\def\cquark    {{\ensuremath{\Pc}}\xspace}

\def\bquark    {{\ensuremath{\Pb}}\xspace}

\def\pion   {{\ensuremath{\Ppi}}\xspace}
\def\piz    {{\ensuremath{\pion^0}}\xspace}

\def\pip    {{\ensuremath{\pion^+}}\xspace}
\def\pim    {{\ensuremath{\pion^-}}\xspace}

\def\kaon    {{\ensuremath{\PK}}\xspace}
  \def\Kbar    {{\kern 0.2em\overline{\kern -0.2em \PK}{}}\xspace}

\def\KorKbar    {\kern 0.18em\optbar{\kern -0.18em K}{}\xspace}

\def\Kzb     {{\ensuremath{\Kbar{}^0}}\xspace}
\def\Kp      {{\ensuremath{\kaon^+}}\xspace}

\def\KS      {{\ensuremath{\kaon^0_{\mathrm{ \scriptscriptstyle S}}}}\xspace}
\def\KL      {{\ensuremath{\kaon^0_{\mathrm{ \scriptscriptstyle L}}}}\xspace}
\def\Kstarz  {{\ensuremath{\kaon^{*0}}}\xspace}

  \def\Dbar    {{\kern 0.2em\overline{\kern -0.2em \PD}{}}\xspace}

\def\DorDbar    {\kern 0.18em\optbar{\kern -0.18em D}{}\xspace}

\def\B       {{\ensuremath{\PB}}\xspace}
\def\Bbar    {{\ensuremath{\kern 0.18em\overline{\kern -0.18em \PB}{}}}\xspace}

\def\BorBbar    {\kern 0.18em\optbar{\kern -0.18em B}{}\xspace}

\def\Bu      {{\ensuremath{\B^+}}\xspace}

\def\jpsi     {{\ensuremath{{\PJ\mskip -3mu/\mskip -2mu\Ppsi\mskip 2mu}}}\xspace}

  \def\Y#1S{\ensuremath{\PUpsilon{(#1S)}}\xspace}% no space before {...}!

\def\proton      {{\ensuremath{\Pp}}\xspace}

\def\Lz          {{\ensuremath{\PLambda}}\xspace}
\def\Lbar        {{\ensuremath{\kern 0.1em\overline{\kern -0.1em\PLambda}}}\xspace}
\def\LorLbar    {\kern 0.18em\optbar{\kern -0.18em \PLambda}{}\xspace}

\def\BF         {{\ensuremath{\mathcal{B}}}\xspace}

\def\BR         {\BF}
\newcommand{\decay}[2]{\ensuremath{#1\!\to #2}\xspace}         % {\Pa}{\Pb \Pc}

\def\to                 {\ensuremath{\rightarrow}\xspace}

\def\CP                {{\ensuremath{C\!P}}\xspace}

\def\AT#1     {\ensuremath{A_{\mathrm{T}}^{#1}}\xspace}           % 2

\def\C#1      {\ensuremath{\mathcal{C}_{#1}}\xspace}                       % 9
\def\Cp#1     {\ensuremath{\mathcal{C}_{#1}^{'}}\xspace}                    % 7
\def\Ceff#1   {\ensuremath{\mathcal{C}_{#1}^{\mathrm{(eff)}}}\xspace}        % 9  
\def\Cpeff#1  {\ensuremath{\mathcal{C}_{#1}^{'\mathrm{(eff)}}}\xspace}       % 7
\def\Ope#1    {\ensuremath{\mathcal{O}_{#1}}\xspace}                       % 2
\def\Opep#1   {\ensuremath{\mathcal{O}_{#1}^{'}}\xspace}                    % 7

\newcommand{\tev}{\ifthenelse{\boolean{inbibliography}}{\ensuremath{~T\kern -0.05em eV}\xspace}{\ensuremath{\mathrm{\,Te\kern -0.1em V}}}\xspace}
\newcommand{\gev}{\ensuremath{\mathrm{\,Ge\kern -0.1em V}}\xspace}
\newcommand{\mev}{\ensuremath{\mathrm{\,Me\kern -0.1em V}}\xspace}
\newcommand{\kev}{\ensuremath{\mathrm{\,ke\kern -0.1em V}}\xspace}
\newcommand{\ev}{\ensuremath{\mathrm{\,e\kern -0.1em V}}\xspace}
\newcommand{\gevc}{\ensuremath{{\mathrm{\,Ge\kern -0.1em V\!/}c}}\xspace}
\newcommand{\mevc}{\ensuremath{{\mathrm{\,Me\kern -0.1em V\!/}c}}\xspace}
\newcommand{\gevcc}{\ensuremath{{\mathrm{\,Ge\kern -0.1em V\!/}c^2}}\xspace}
\newcommand{\gevgevcccc}{\ensuremath{{\mathrm{\,Ge\kern -0.1em V^2\!/}c^4}}\xspace}
\newcommand{\mevcc}{\ensuremath{{\mathrm{\,Me\kern -0.1em V\!/}c^2}}\xspace}

\def\mm   {\ensuremath{\mathrm{ \,mm}}\xspace}

\def\mum  {\ensuremath{{\,\upmu\mathrm{m}}}\xspace}

\def\invfb   {\ensuremath{\mbox{\,fb}^{-1}}\xspace}

\newcommand{\chisq}{\ensuremath{\chi^2}\xspace}

\def\gsim{{~\raise.15em\hbox{$>$}\kern-.85em
          \lower.35em\hbox{$\sim$}~}\xspace}
\def\lsim{{~\raise.15em\hbox{$<$}\kern-.85em
          \lower.35em\hbox{$\sim$}~}\xspace}

\def\ptot       {\mbox{$p$}\xspace}
\def\pt         {\mbox{$p_{\mathrm{ T}}$}\xspace}

\def\evtgen     {\mbox{\textsc{EvtGen}}\xspace}

\def\geant      {\mbox{\textsc{Geant4}}\xspace}

\def\photos     {\mbox{\textsc{Photos}}\xspace}

\def\pythia     {\mbox{\textsc{Pythia}}\xspace}

\def\tell1  {TELL1\xspace}
\def\ukl1   {UKL1\xspace}

%% file: our-symbols-def.tex
\newcommand{\BDTcb}{\ensuremath{\text{BDT}_{\rm cb}}\xspace}
\newcommand{\BDTmu}{\ensuremath{\text{BDT}_\mu}\xspace}

\newcommand{\pl}{\ensuremath{p_L}\xspace}

\def\BuToJPsiK {\decay{\Bu}{\jpsi\Kp}}

\newcommand{\Ks}{\KS}
\newcommand{\Kl}{\KL}

\newcommand{\Ksmumu}{\ensuremath{\Ks\to\mu^+\mu^-}\xspace}

\newcommand{\Klmumu}{\ensuremath{\Kl\to\mu^+\mu^-}\xspace}
\newcommand{\Kspipi}{\ensuremath{\Ks\to\pi^+\pi^-}\xspace}

\newcommand{\Kspimunu}{\ensuremath{\Ks\to\pi^+\mu^-\bar{\nu}_\mu}\xspace}

\newcommand{\BuJpsiK}{\ensuremath{B^+\to J/\psi K^+}\xspace}

\newcommand{\BRof}[1]{\ensuremath{{\cal B}(#1)}\xspace}

\newcommand{\epsmuid}{\ensuremath{\epsilon_{\mu {\rm{ID}}}}\xspace}

\newcommand{\swave}{{S-wave}\xspace}

\newcommand{\tosm}{TOS\ensuremath{_{\mu}}\xspace}
\newcommand{\tosmm}{TOS\ensuremath{_{\mu\mu}}\xspace}

%% file: title-LHCb-PAPER.tex
% $Id: title-LHCb-PAPER.tex 95682 2016-07-21 12:13:58Z michaelt $
% ===============================================================================
% Purpose: LHCb-PAPER journal paper title page template
% Author: 
% Created on: 2010-09-25
% ===============================================================================

%%%%%%%%%%%%%%%%%%%%%%%%%
%%%%%  TITLE PAGE  %%%%%%
%%%%%%%%%%%%%%%%%%%%%%%%%
\begin{titlepage}
\pagenumbering{roman}

% Header ---------------------------------------------------
\vspace*{-1.5cm}
\centerline{\large EUROPEAN ORGANIZATION FOR NUCLEAR RESEARCH (CERN)}
\vspace*{1.5cm}
\noindent
\begin{tabular*}{\linewidth}{lc@{\extracolsep{\fill}}r@{\extracolsep{0pt}}}
\ifthenelse{\boolean{pdflatex}}% Logo format choice
{\vspace*{-3.0cm}\mbox{\!\!\!\includegraphics[width=.14\textwidth]{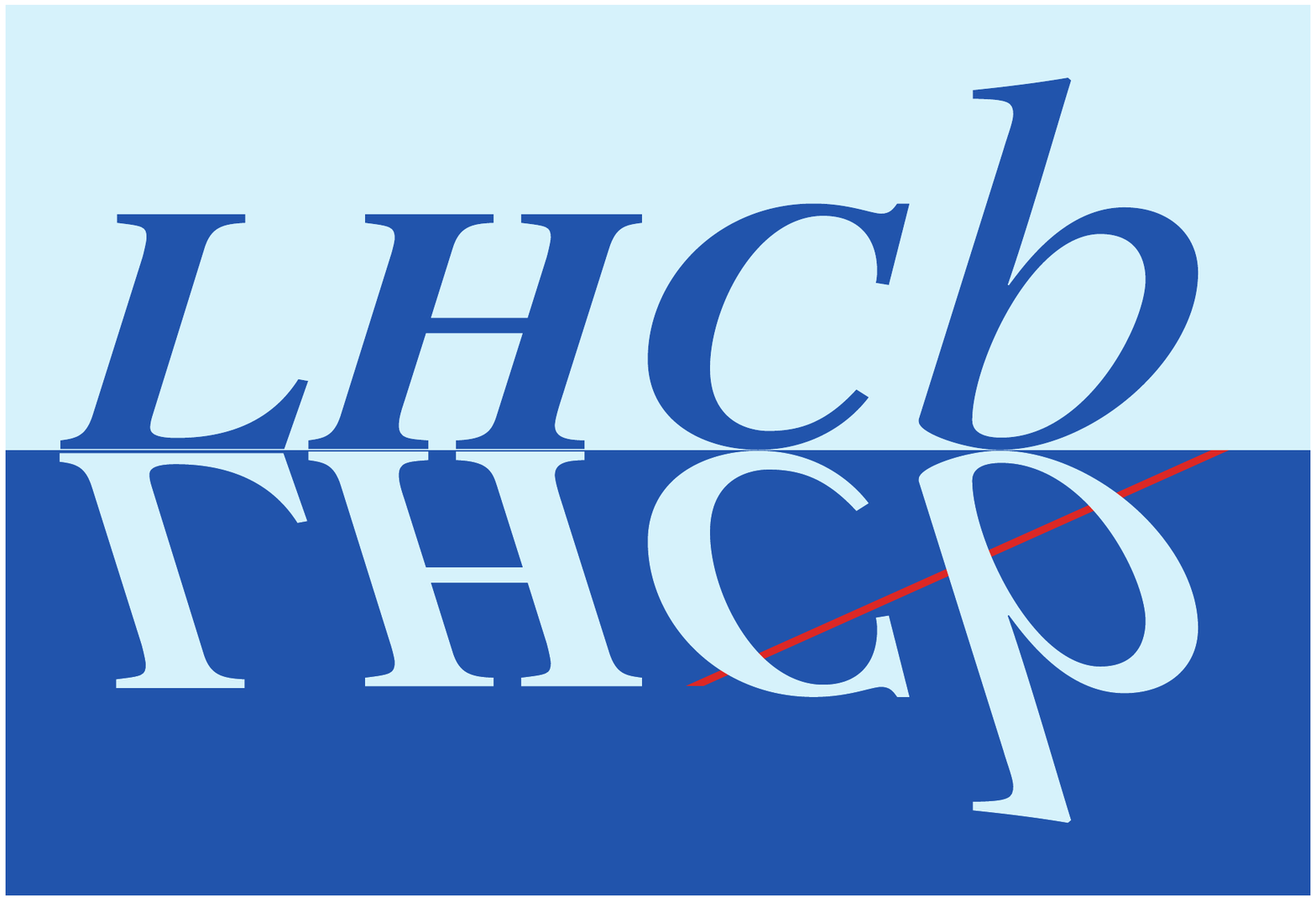}} & &}%
{\vspace*{-1.2cm}\mbox{\!\!\!\includegraphics[width=.12\textwidth]{lhcb-logo.eps}} & &}%
\\
 & & CERN-EP-2017-101 \\  % ID 
 & & LHCb-PAPER-2017-009 \\  % ID 
 & & June 2, 2017 \\
\end{tabular*}

\vspace*{4.0cm}

% Title --------------------------------------------------
{\normalfont\bfseries\boldmath\huge
\begin{center}
  Improved limit on the branching fraction of the rare decay \Ksmumu%using 2012 data
\end{center}
}

\vspace*{2.0cm}

% Authors -------------------------------------------------
\begin{center}
%In the footnote, replace 'paper' by 'Letter' in case of submission to PRL or PLB 
The LHCb collaboration\footnote{Authors are listed at the end of this paper.}
\end{center}

\vspace{\fill}

% Abstract -----------------------------------------------
\begin{abstract}
  \noindent
  A search for the decay \Ksmumu is performed, based on a data sample of proton-proton collisions corresponding to an integrated luminosity of $3\,\invfb$, collected by the LHCb experiment at centre-of-mass energies of 7~and~8\tev. 
The observed yield is consistent with the background-only hypothesis, yielding a limit on the branching fraction of $\BRof{\Ksmumu} < 0.8~(1.0) \times10^{-9}$ at $90\%~(95\%)$ confidence level.
This result improves the previous upper limit on the branching fraction by an order of magnitude.
\end{abstract}

\vspace*{2.0cm}

\begin{center}
  Published in Eur.~Phys.~J.~C
\end{center}

\vspace{\fill}

{\footnotesize 
\centerline{\copyright~CERN on behalf of the \lhcb collaboration, licence \href{http://creativecommons.org/licenses/by/4.0/}{CC-BY-4.0}.}}
\vspace*{2mm}

\end{titlepage}

%%%%%%%%%%%%%%%%%%%%%%%%%%%%%%%%
%%%%%  EOD OF TITLE PAGE  %%%%%%
%%%%%%%%%%%%%%%%%%%%%%%%%%%%%%%%

%  empty page follows the title page ----
\newpage
\setcounter{page}{2}
\mbox{~}
%\newpage
%
%% Author List ----------------------------
%%  You need to get a new author list!
%\input{LHCb_authorlist.tex}
%
%The author list for journal publications is provided by the Membership Committee shortly after 'approval to go to paper' has been given.
%%It will be made available on the page
%%\verb!http://www.physik.uzh.ch/~strauman/forMemCo/LHCb-PAPER-XXXX-XXX/! .
%It will be sent to you by email shortly after a paper number has beens assigned.
%The author list should be included already at first circulation, 
%to allow new members of the collaboration to verify whether they have been included correctly.
%Occasionally a misspelled name is corrected or associated institutions become full members.
%In that case, a new author list will be sent to you.
%In case line numbering doesn't work well after including the authorlist, try moving the \verb!\bigskip! after the last author to a separate line.
%
%
%The authorship for Conference Reports should be ``The LHCb
%  collaboration'', with a footnote giving the name(s) of the contact
%  author(s), but without the full list of collaboration names.

\cleardoublepage

%% file: introduction.tex
\section{Introduction}
\label{sec:Introduction}

In the Standard Model (SM), the unobserved \Ksmumu decay proceeds only through a Flavour-Changing Neutral Current (FCNC) transition, which cannot occur at tree level.
It is further suppressed by the small amount of \CP violation in kaon decays, since  the \swave component of the decay is forbidden when \CP is conserved. 
In the SM, the decay amplitude is expected to be dominated by long distance contributions, which can be constrained using the observed decays \decay{\KS}{\gamma\gamma} and \decay{\KL}{\piz\gamma\gamma}, leading to the prediction for the branching fraction $\BRof\Ksmumu = (5.0 \pm 1.5) \times 10^{-12}$~\cite{Ecker:1991ru,Isidori:2003ts}.
The predicted branching fraction for the \KL decay is \mbox{$(6.85 \pm 0.32) \times 10^{-9}$}~\cite{DAmbrosio:1994fgc},  in excellent agreement with the experimental world average $\BR(\decay{\KL}{\mumu}) = (6.84 \pm 0.11) \times 10^{-9}$~\cite{PDG2016}.
The prediction for \Ksmumu is currently being updated with a dispersive treatment, which
leads to sizeable corrections in other \KS leptonic decays~\cite{Lewis}.

Due to its suppression in the SM, the \Ksmumu decay is sensitive to possible contributions from dynamics beyond the SM, notably from light scalars with \CP-violating Yukawa couplings\cite{Ecker:1991ru}. 
Contributions up to one order of magnitude above the SM branching fraction expectation naturally arise in many models and are compatible with the present bounds from other FCNC processes~\cite{Isidori:2003ts}.
An upper limit on \BRof\Ksmumu close to $10^{-11}$ could be translated into model-independent bounds on the \CP-violating phase of the \decay{s}{d\ellp\ellm} amplitude. This would be very useful to discriminate between scenarios beyond the SM if other modes, such as \decay{\Kp}{\pip\neu\neub}, indicate a non-SM enhancement.%~\cite{Isidori:2003ts}.

The current experimental limit, $\BR(\Ksmumu) < 9 \times 10^{-9}$ at $90\%$ confidence level (CL), was obtained using $pp$ collision data corresponding to $1.0\,\invfb$ of integrated luminosity at a centre-of-mass energy $\sqrt{s}=7~\tev$, collected with the \lhcb detector in 2011~\cite{LHCb-PAPER-2012-023}.
This result improved the previous upper limit~\cite{Gjesdal:880748} but is still three orders of magnitude above the predicted SM level.

In this paper, an update of the search for the \Ksmumu decay is reported.
Its branching fraction is measured using the known \Kspipi decay as normalisation.
The analysis is performed on a data sample corresponding to $2\,\invfb$ of integrated luminosity at $\sqrt{s}=8~\tev$, collected in 2012, and the result is combined with that from the previous \lhcb analysis~\cite{LHCb-PAPER-2012-023}.
Besides the gain in statistical precision due to the larger data sample, the sensitivity is noticeably increased with respect to the previous result due to a higher trigger efficiency, as well as other improvements to the analysis that are discussed in the following sections.

An overview on how \Ksmumu decays are detected and triggered in \lhcb is given in Sect.~\ref{sec:kaonlhcb}, while the strategy for this measurement is outlined in Sect.~\ref{sec:strategy}.
Details of background suppression and the resulting sensitivity are given in Sects.~\ref{sec:backgrounds} and~\ref{sec:sensitivity}, respectively.
The final result, taking into account the systematic uncertainties discussed in Sect.~\ref{sec:systematics}, is given in Sect.~\ref{sec:results}.

%% file: detectoranddata.tex
\section{\boldmath\KS decays in \lhcb}

\label{sec:Detector}
\label{sec:kaonlhcb}

The \lhcb detector~\cite{Alves:2008zz,LHCb-DP-2014-002} is a single-arm forward spectrometer covering the \mbox{pseudorapidity} range $2<\eta <5$, designed for the study of particles containing \bquark or \cquark quarks.
The detector includes a high-precision tracking system consisting of a silicon-strip vertex locator (VELO) surrounding the $pp$ interaction region%~\cite{LHCb-DP-2014-001}
, a large-area silicon-strip detector located upstream of a dipole magnet with a bending power of about $4{\mathrm{\,Tm}}$, and three stations of silicon-strip detectors and straw drift tubes %~\cite{LHCb-DP-2013-003}
placed downstream of the magnet.
The tracking system provides a measurement of momentum, \ptot, of charged particles with a relative uncertainty that varies from $0.5\%$ at low momentum to $1.0\%$ at $200\,\gevc$.
The minimum distance of a track to a primary vertex (PV), the impact parameter (IP), is measured with a resolution of $(15+29/\pt)\,\mum$, where \pt is the component of the momentum transverse to the beam, in \gevc.
Different types of charged hadrons are distinguished using information from two ring-imaging Cherenkov detectors%~\cite{LHCb-DP-2012-003}
~(RICH). 
Photons, electrons and hadrons are identified by a calorimeter system consisting of scintillating-pad and preshower detectors, an electromagnetic calorimeter and a hadronic calorimeter. Muons are identified by five stations which alternate layers of iron and multiwire proportional chambers%~\cite{LHCb-DP-2012-002}.

The online event selection is performed by a trigger~\cite{LHCb-DP-2012-004}, which consists of a hardware stage, based on information from the calorimeter and muon systems, followed by a two-step software stage, which applies a full event reconstruction.
Candidates are subsequently classified as TOS, if the event is triggered on the signal candidate, or TIS, if triggered by other activities in the detector, independently of signal.
Only candidates that are classified as TOS at each trigger stage are used to search for \Ksmumu decays. 

The trigger selection constitutes the main limitation to the efficiency for detecting \Ks decays.
A muon is only selected at the hardware stage when it is detected in all muon stations, implying a momentum larger than about $5\,\gevc$, and a \pt above $1.76\,\gevc$.
These thresholds have an efficiency of order $1\%$ for \Ksmumu decays.

In the first step of the software trigger, all charged particles with $\pt>500\,\mevc$ are reconstructed.
At this stage most signal decays are triggered either by requiring a reconstructed track loosely identified as a muon~\cite{LHCb-DP-2012-004,LHCb-DP-2013-001}, with $\text{IP}>0.1\,\mm$ and $\pt>1.0\,\gevc$, or by finding two oppositely charged muon candidates forming a detached secondary vertex (SV).
Since these two categories, hereafter referred to as \tosm and \tosmm, induce different kinematic biases on the signal and background candidates, the analysis steps described below are performed independently on each category. The two categories are made mutually exclusive by applying the \tosmm selection only to candidates not already selected by \tosm.

In the second software trigger stage, an offline-quality event reconstruction is performed.
Signal candidates are selected requiring a dimuon with $\pt>600\,\mevc$ detached from the primary vertex, with both tracks having $\pt>300\,\mevc$. 
In the 2011 data taking, the dimuon mass was required to be larger than $1\,\gevcc$ in the second software trigger stage. This excluded the \KS region, making the use of TIS candidates necessary.
Due to the trigger reoptimisation, no mass requirements were applied during 2012 and a lower \pt threshold for reconstructed tracks was used. 
According to simulation, these changes improve the trigger efficiency over the previous analysis~\cite{LHCb-PAPER-2012-023} by about a factor $2.5$.

Due to its large and well-known branching fraction and its similar topology, the \Kspipi decay is taken as the normalisation mode.
A large sample of candidates is obtained from an unbiased trigger, which does not apply any selection requirement.

Despite the low trigger efficiency, the study detailed in this paper profits from the unprecedented number of \KS produced at the LHC, $\mathcal{O}(10^{13})$ per \invfb of integrated luminosity within the \lhcb acceptance, and from the fact that about $40\%$ of these \KS decays occur inside the \velo region.
For such decays, the \Ks invariant mass is reconstructed with a resolution of about $4\,\mevcc$.

The analysis makes use of large samples of simulated collisions containing a signal decay, or background decays which can be reconstructed as the signal, and contaminate the $\mu\mu$ invariant mass distribution, such as \Kspipi or \Kspimunu.\footnote{The inclusion of charge-conjugate processes is implied throughout.}
In the simulation, $pp$ collisions are generated using \pythia~\cite{Sjostrand:2006za,*Sjostrand:2007gs} with a specific \lhcb configuration~\cite{LHCb-PROC-2010-056}. 
Decays of hadronic particles are described by \evtgen~\cite{Lange:2001uf}, in which final-state radiation is generated using \photos~\cite{Golonka:2005pn}.
The interaction of the generated particles with the detector, and its response, are implemented using the \geant toolkit~\cite{Allison:2006ve, *Agostinelli:2002hh} as described in Ref.~\cite{LHCb-PROC-2011-006}.

%% file: strategy.tex
\section{Selection and search strategy}
\label{sec:strategy}

Common offline preselection criteria are applied to \Ksmumu and \Kspipi candidates to cancel many systematic effects in the ratio.
Candidates are required to decay in the \velo region, where the best \Ks mass resolution is achieved.
The two reconstructed tracks must have momentum smaller than $100\,\gevc$ and quality requirements are set on the track and secondary vertex fits.
The SV must be well detached from the PV by requiring the \Ks decay time to be larger than $8.95\,$ps, $10\%$ of the \Ks mean lifetime.
The \Ks IP must be less than $0.4\,\mm$, while the two charged tracks are required to be incompatible with originating from any PV.

Decays of \Lz baryons to $\proton\pim$ are suppressed by removing candidates close to the expected ellipses in the Armenteros-Podolanski (AP) plane~\cite{armenteros}.
In this plane the \pt of the final-state particles under the pion mass hypothesis is plotted versus the longitudinal momentum asymmetry, defined as
$\alpha = (\pl^+ - \pl^-)/(\pl^+ + \pl^-)$, where $\pl^\pm$ is the longitudinal momentum of the charged tracks.
Both \pt and \pl are considered with respect to the direction of the mother particle.
The \Ks decays are symmetrically distributed on the AP plane while \Lz decays produce two ellipses at low \pt and $|\alpha|\sim0.7$.
A kaon veto, based on the response of the RICH detector, is used to suppress \decay{\Kstarz}{\Kp\pim} decays and other possible final states including a charged kaon.

The preselection reduces the combinatorial background, arising from candidates formed from secondary hadronic collisions in the detector material or from spurious reconstructed SV.
The purity of the \Kspipi sample used for normalisation, whose mass distribution is shown in Fig.~\ref{fig:InvMass}, is estimated from a fit to the mass spectrum to be $99.8\%$.
The fraction of events with more than one candidate is less than $0.1\%$ for signal and $4\%$ for the normalisation channel, and all candidates are retained.
Additional discrimination against backgrounds for the signal mode is achieved through the use of two multivariate discriminants. The first is designed to further suppress combinatorial background, and the second to reduce the number of \Kspipi decays in which both pions are misidentified as muons.

After requirements on the output of these discriminants have been applied, the number of signal candidates is obtained by fitting the \Ksmumu mass spectrum. The mass sidebands provide a data-driven estimation of the residual background by extrapolation into the signal region.
The number of candidates is converted into a branching fraction using the yield of the \Kspipi normalisation mode, and the estimated relative efficiency. 
Events in the \Ks mass region are scrutinised only after fixing the analysis strategy.

%% file: backgrounds.tex
\section{Backgrounds}
\label{sec:backgrounds}

\begin{figure}[tb]
\centering
\centerline{\includegraphics[width=.85\textwidth]{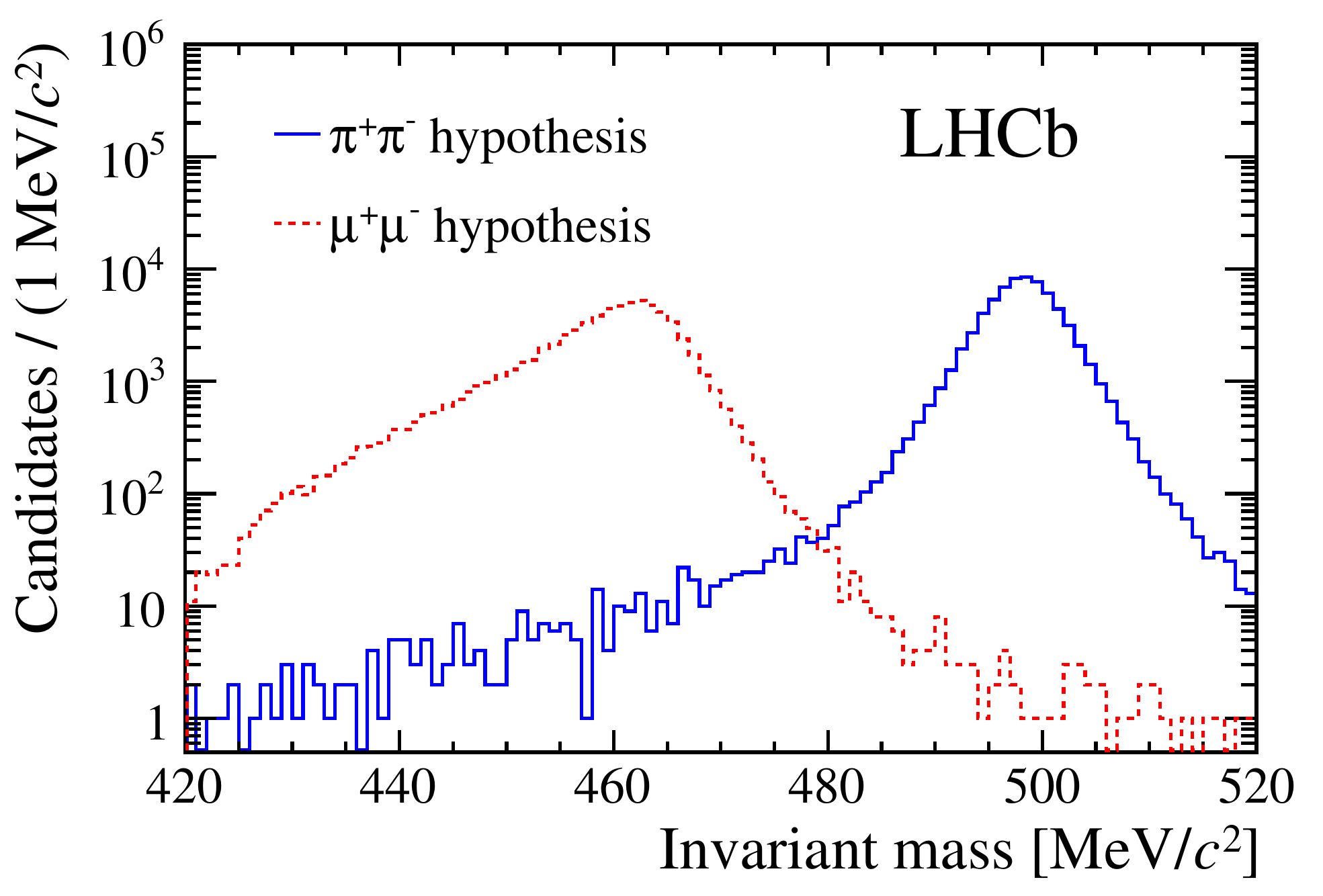}}
\caption{Reconstructed mass for \Kspipi decays in trigger-unbiased events, computed assuming the muon (dashed red line) or pion (solid blue line) mass for the final-state tracks. Candidates satisfy the selection criteria described in the text.}
\label{fig:InvMass}
\end{figure}

The \Ksmumu sample contains two main sources of background. Combinatorial background candidates are expected to exhibit a smooth mass distribution, and can therefore be estimated from the sidebands.
The other relevant source of background is due to \Kspipi decays where both pions pass the loose muon identification requirements after the trigger stage.
This can be due either to \decay{\pip}{\mup\neum} decays or to random association of muon detector hits with the pion trajectory.
In such cases the \Ks mass, reconstructed with a wrong mass hypothesis for the final-state particles, is underestimated by $39\,\mevcc$ on average, as shown in Fig.~\ref{fig:InvMass}.
Despite the excellent mass resolution, the right-hand tail of the   
reconstructed mass distribution under the dimuon hypothesis extends into the \KS signal mass range and, given   
the large branching fraction of the \Kspipi mode, constitutes a
nonnegligible background.
Two multivariate discriminants, based on a Boosted Decision Tree (BDT) algorithm~\cite{Breiman,AdaBoost}, are applied on the preselected candidates to improve the signal discrimination with respect to these backgrounds. 

The first discriminant, named hereafter \BDTcb, aims to reduce the combinatorial background, exploiting the different decay topologies, kinematic spectra and reconstruction qualities of signal and combinatorial candidates.
It is optimised separately for each trigger category.
A set of ten input variables is used in \BDTcb: the \Ks \pt and IP, the minimum IP of the two charged tracks, the angle between the positively charged final-state particle and the \Ks flight direction in the \Ks rest frame, the \chisq of the SV fit, the distance of closest approach between the two tracks, an SV isolation variable, defined as the difference in vertex-fit \chisq when the next nearest track is included in the vertex fit, and the SV absolute position coordinates.  
The SV position is particularly important, since a large fraction of the background is found to originate from interactions in the detector material.
This set of variables does not distinguish between \Ksmumu and \Kspipi decays as it does not contain quantities related to muon identification and ignores the \Ks candidate invariant mass distribution.

The signal training sample for \BDTcb is composed of \Ksmumu simulated candidates passing the trigger and preselection criteria. 
A signal training sample consisting of \Kspipi decays in data is also used as a cross-check, as explained in Sect.~\ref{sec:systematics}.
The background training sample is made from \Ksmumu data candidates surviving the trigger and preselection requirements with reconstructed mass in the range $[520,600]\,\mevcc$.
Since candidates in the same mass region are also used to estimate the residual background, the training is performed using 
 a {\it k-fold} cross-validation technique~\cite{Blum:1999:BHB:307400.307439} to avoid any possible effect of overtraining.

A loose requirement on the \BDTcb output is applied to suppress the combinatorial background.
The cut is chosen to remove 99\% of the background training candidates.
The corresponding signal efficiency is about $56\%$ and $66\%$ for the \tosm and \tosmm trigger categories, respectively. 
To exploit further the information provided by the discriminant, the candidates surviving this requirement are allocated to ten bins according to their \BDTcb value, with bounds defined in order to have approximately equal population of signal training candidates in each bin. 

The background from misidentified \Kspipi decays is further reduced with the second multivariate discriminant, called \BDTmu. Its input includes the position, time and number of detector hits around the extrapolated track position to each muon detector station, a global match \chisq between the muon hit positions and the track extrapolation,  and other variables related to the tracking and the response of the RICH and calorimeter detectors.

To train the \BDTmu discriminant, highly pure samples of $1.2$ million pions and $0.68$ million muons are obtained from TIS-triggered \Kspipi and \BuJpsiK decays, respectively.
In the latter case, a probe muon from the \jpsi is required to be TIS at all trigger stages, while stringent muon identification requirements are set on the other muon, reaching an estimated purity for muons above $99.9\%$.
Before using it in the \BDTmu training, the muon sample is weighted to have the same two-dimensional distribution in \ptot and \pt as the pion sample, as well as the same distribution of number of tracks in the event.
This is to prevent the \BDTmu from discriminating pions and muons using these variables, which are included in the input because of their strong correlation with the identification variables. 
Weighting also allows optimisation of the discrimination power for the kinematic spectrum relevant to this search.

The level of misidentification of the discriminant for a pion from \Kspipi decay is found to be $0.4\%$ for $90\%$ muon efficiency.
This reduces the level of double misidentification background, for a given efficiency, by about a factor of four with respect to the discriminant used in the previous publication~\cite{LHCb-PAPER-2012-023}, which was not tuned specifically for \Ksmumu searches.

The \BDTmu discriminant is trained using half of the \BuJpsiK sample, while the other half is used to evaluate the muon identification 
efficiency as a function of (\ptot, \pt).  These values are used to compute the efficiency of a \BDTmu requirement on 
the candidate \Ksmumu decays after selection and trigger requirements, in each bin of the \BDTcb discriminant. 
The muon spectra assumed in this calculation are obtained from simulated decays, weighted to better reproduce the \Ks \pt spectrum 
observed in \Kspipi candidates.

The \BDTmu requirement on the signal candidates is optimised by maximising the figure of merit~\cite{Punzi:2003bu} $\epsmuid/(\sqrt{N_{\rm bg}}+a/2)$, with $a=3$, where $\epsmuid$ is the signal efficiency and $N_{\rm bg}$ the expected background yield.  %studying the dependence of the signal efficiency \epsmuid and expected background $B$ as a function of the \BDTmu threshold value.
The latter is estimated from a fit to the mass distribution, after removing candidates in the range $[492,504]\,\mevcc$ around the \Ks mass,
and extrapolating the result into the signal region.
This optimisation is performed independently for the two trigger categories, with no significant difference found as a function of the \BDTcb bin. The optimal threshold corresponds to a signal efficiency of $\epsmuid \sim 98\%$ in both cases. 

Other possible sources of background have been explored and found to give negligible contribution to this search.
The irreducible background due to \decay{\KL}{\mumu} decays and from \KS--\KL interference is evaluated from the known \Klmumu branching fraction and lifetime, and by studying the decay-time dependence of the selection efficiency for \Kspipi decays in data.
The yield from this background becomes comparable to the signal for a branching fraction lower than $2 \times 10^{-11}$, which is well below the sensitivity of this search.

Semileptonic \decay{\Kzb}{\pip\mun\neumb} decays with pion misidentification provide another possible source of background.
Simulated events, where the pion is forced to decay to $\mu\nu$ within the detector, are used to determine the efficiency of the offline selection requirements. No event survives the trigger selection.
Under the very conservative hypothesis that the trigger efficiency is the same as in \Ksmumu decays, the expected yields from both \KL and \KS semileptonic decays are negligible.

Decays including a dimuon from resonances, like \decay{\omega}{\piz\mumu} and \decay{\eta}{\mumu\gamma}, do not produce peaking structures in the mass distribution, and are accounted for in the combinatorial background.

%% file: sensitivity.tex
\section{Search sensitivity}
\label{sec:sensitivity}

The observed number of \Ksmumu candidates is converted into a branching fraction using the normalisation mode and its precisely known branching fraction  $\BRof\Kspipi = 0.6920\pm0.0005 $ \cite{PDG2016}.
The computation is made in every \BDTcb bin $i$ and trigger category $j$ as follows
\begin{equation}
\BRof\Ksmumu  = \BRof\Kspipi \cdot \frac{
  \epsilon^{\pi\pi}}{\epsilon^{\mu\mu}_{ij}}
\cdot\frac{N^{\mu\mu}_{ij}}{N^{\pi\pi}}
 \equiv \alpha_{ij} N^{\mu\mu}_{ij},
\end{equation}
where $N^{\mu\mu}_{ij}$ and $N^{\pi\pi}$ denote the background-subtracted yields for the signal and normalisation
modes, respectively.
The total selection efficiencies $\epsilon$ can be factorised as
\begin{equation}
  \frac{\epsilon^{\pi\pi}}{\epsilon^{\mu\mu}_{ij}} =
  \frac{\epsilon^{\pi\pi}_{\rm{sel}}}{\epsilon^{\mu\mu}_{\rm{sel}}} \times
  \frac{\epsilon^{\pi\pi}_{{\rm{trig}}}}{\epsilon^{\mu\mu}_{{\rm{trig}};j}} \times
  \frac{1}{\epsilon^{\mu\mu}_{{\text{BDT}};ij}} \times
  \frac{1}{\epsilon_{\mu {\rm{ID}};ij}}.
\end{equation}

The first factor refers to the offline selection requirements, which are applied identically to both modes and cancel to first order in the ratio;
the residual difference is mainly due to the different interaction cross-sections for pions and muons with the detector material, and is estimated from simulation. 
The second factor is the ratio of trigger efficiencies;
the efficiency for the signal is determined from simulation, with its systematic uncertainty estimated from data-driven checks, while that for the normalisation mode is the efficiency of the random trigger used to select \Kspipi, $(9.38 \pm 1.01)\times 10^{-8}$.
The third factor reflects the fraction of candidates in each \BDTcb bin, and is also determined from simulation.
Finally, the efficiency of the \BDTmu requirement is obtained from the \BuJpsiK calibration sample described in Sect.~\ref{sec:backgrounds}, for each \BDTcb bin and trigger category.

To account for the difference between the kaon \pt spectra observed in the \Kspipi decays in data and simulation, all efficiencies obtained from simulation are computed 
in six roughly equally populated \pt bins. A weighted average of the efficiencies is then performed, where the weights
are determined from the yields in each bin observed in data for \Kspipi candidates.

The resulting values for the single candidate sensitivity $\alpha_{ij}$ are reported in Table~\ref{tab:alpha}. 
The quoted uncertainties are statistical only. They are separated between the uncertainty on $\epsilon^{\mu\mu}_{{\text{BDT}};ij}$, due to the limited statistics of simulated data and uncorrelated among \BDTcb bins, and all the other statistical uncertainties, which are conservatively considered as fully correlated among bins within the same trigger category.
Table~\ref{tab:alpha} also presents the number of candidates after the inspection of the signal region.
The separation between signal and background is presented in Sect.~\ref{sec:results}.

\begin{table}[tbp]
\caption{Values of the single candidate sensitivity $\alpha_{ij}$ and the number of candidates $N^{K}_{ij}$ compatible with the \Ks mass (reconstructed mass
  in the range $[492,504]\,\mevcc$), for each \BDTcb bin $i$ and trigger category $j$. Only statistical uncertainties are given. The first uncertainty is uncorrelated, while the second is fully correlated among the \BDTcb bins of the same trigger category.
  %The $N^{K}_{ij}$ yields are expected to be dominated by background candidates.
}
\label{tab:alpha}
\begin{center}
    \begin{tabular}{ccccc}
    \toprule
    Bin $i$  & $\alpha_{i \text{TOS}_\mu} (\times 10^{-10})$ & $\alpha_{i \text{TOS}_{\mu\mu}} (\times 10^{-9})$ & $N^{K}_{i \text{TOS}_\mu}$ & $N^{K}_{i \text{TOS}_{\mu\mu}}$\\ \midrule
    1 & $7.48 \pm  0.84 \pm  0.16$          & $5.30 \pm  0.72 \pm  0.12$ & 49 & 13 \\
    2 & $7.72 \pm  0.87 \pm  0.17$          & $4.71 \pm  0.63 \pm  0.10$ & 28 & 9  \\
    3 & $7.85 \pm  0.89 \pm  0.18$          & $4.88 \pm  0.65 \pm  0.11$ & 9  & 14 \\
    4 & $7.93 \pm  0.89 \pm  0.19$          & $4.66 \pm  0.62 \pm  0.10$ & 18 & 10 \\
    5 & $7.53 \pm  0.85 \pm  0.18$          & $4.65 \pm  0.61 \pm  0.10$ & 6  &  3 \\
    6 & $7.78 \pm  0.88 \pm  0.19$          & $4.95 \pm  0.66 \pm  0.11$ & 2  &  2 \\
    7 & $7.56 \pm  0.85 \pm  0.19$          & $4.60 \pm  0.61 \pm  0.10$ & 3  &  1 \\
    8 & $7.90 \pm  0.89 \pm  0.19$          & $5.00 \pm  0.67 \pm  0.11$ & 2  &  1 \\
    9 & $7.81 \pm  0.88 \pm  0.18$          & $4.72 \pm  0.63 \pm  0.11$ & 1  &  1 \\
    10 & $7.75 \pm  0.87 \pm  0.17$         & $4.66 \pm  0.62 \pm  0.11$ & 0  &  0 \\
    \bottomrule
    \end{tabular}
\end{center}
\end{table}

%% file: systematics.tex
\section{Systematic uncertainties}
\label{sec:systematics}

Several systematic effects, summarised in Table~\ref{tab:syst}, contribute to the uncertainty on the normalisation factors. 
Tracking efficiencies are not perfectly reproduced in simulated events.
Corrections based on a \decay{\jpsi}{\mumu} data control sample are determined as a function of the muon $\ptot$ and $\eta$.
The average effect of these corrections on the ratio $\epsilon^{\pi\pi}_{\rm sel}/\epsilon^{\mu\mu}_{\rm sel}$ and its standard deviation, added in quadrature, leads to a systematic uncertainty of $0.4\%$.

The distributions of all variables relevant to the selection are compared in data and simulation for \Kspipi decays.
The largest differences are found in the kaon \pt and its decay vertex radial position. 
The effect on $\epsilon^{\pi\pi}_{\rm sel}/\epsilon^{\mu\mu}_{\rm sel}$ of applying a two-dimensional weight to account for these discrepancies is taken as a systematic uncertainty, and amounts to a relative $1.9\%$ and $1.8\%$ for the \tosm and \tosmm trigger categories, respectively. 

The difference between data and simulation in the kaon \pt spectrum could also affect the other factors in the computation of $\alpha_{ij}$.
An additional uncertainty is assigned by repeating the whole calculation with a finer binning in \pt.
Due to the limited size of the data samples, this is possible only in the \tosm category.
The average relative change in $\alpha_{ij}$, $4.3\%$, is assigned as an uncertainty for the \tosmm category.

A specific cross-check is performed to validate the efficiencies predicted by the simulation for the \BDTcb requirements. 
An alternative discriminant is made using a signal training sample consisting of
trigger-unbiased \Kspipi decays, selected with additional kinematic criteria which mimic the effect
of the muon trigger selections. The distributions of this alternative discriminant in data and simulation are found to agree within the statistical uncertainty, and no systematic uncertainty is assigned.
 
The uncertainty due to the simulation of TOS selections in the first two trigger stages is assessed by comparing the trigger efficiency in simulation and data, using a control sample of \BuToJPsiK decays.   
The resulting relative differences, $8.1\%$ for \tosm and $11.5\%$ for \tosmm, are assigned as systematic uncertainties.
No uncertainty is considered for the selection in the last trigger stage, which is based on the same offline kinematic variables used in the selection, for which a systematic  uncertainty is already assigned.

The uncertainty on $\epsilon_{\mu {\rm{ID}};ij}$ is estimated from half the difference between the values obtained with and without the weighting of the \BuToJPsiK sample used in the determination of the muon identification efficiency.
This results in an uncertainty of 0.2\% and 0.3\% for the \tosm and \tosmm categories, respectively, which is comparable to the statistical uncertainties on these efficiencies due to the limited size of the \BuToJPsiK samples.

Systematic uncertainties on the signal yields $N^{\mu\mu}_{ij}$ are related to the assumed models for the reconstructed \Ks mass distribution, determined from simulation.
Possible discrepancies from the shape in data are estimated by comparing the shape of the invariant mass distribution in data and simulation for \Kspipi decays, leading to a relative $0.8\%$ systematic uncertainty on the signal yield.
The final fit for the determination of the branching fraction is performed with two different background models, as discussed in Sect.~\ref{sec:results}. This leads to a relative variation on the branching fraction of 0.9\%, which is assigned as a systematic uncertainty.

\begin{table}[tpb]
    \begin{center}
    \caption{Relevant systematic uncertainties on the branching fraction.
    They are separated, using horizontal lines, into relative uncertainties on (i) $\alpha_{ij}$, (ii) on the signal yield from the signal model used in the mass fit, and (iii) on the branching fraction, obtained combining the two categories, from the background model.\label{tab:syst}}
    \begin{tabular}{lrr}
    \toprule
    Source              & \tosm                                                        & \tosmm  \\ \midrule
                      %& \multicolumn{2}{c}{Uncertainties on normalisation factor} \\
    Tracking            & $0.4\%$                                                        & $0.4\%$   \\
    Selection           & $1.9\%$                                                        & $1.8\%$   \\
    Trigger             & $8.1\%$                                                          & $11.5\%$    \\
    \KS \pt spectrum    & $4.3\%$                                                        & $4.3\%$   \\
    Muon identification & $0.2\%$                                                        & $0.3\%$   \\ \midrule
                      %& \multicolumn{2}{c}{Uncertainties on signal yield} \\
    Signal mass shape   & $0.8\%$                                                        & $0.8\%$   \\\midrule
                      %& \multicolumn{2}{c}{Uncertainties on branching fraction} \\
    Background shape    & \multicolumn{2}{c}{$0.9\%$}  \\
    \bottomrule
    \end{tabular}
    \end{center}
\end{table}

%% file: results.tex
\section{Results}
\label{sec:results}

\begin{figure}[t]
    \begin{center}
        %\centerline{\includegraphics[width=.85\textwidth]{figs/fits.pdf}}
        \subfloat{\includegraphics[scale=0.4]{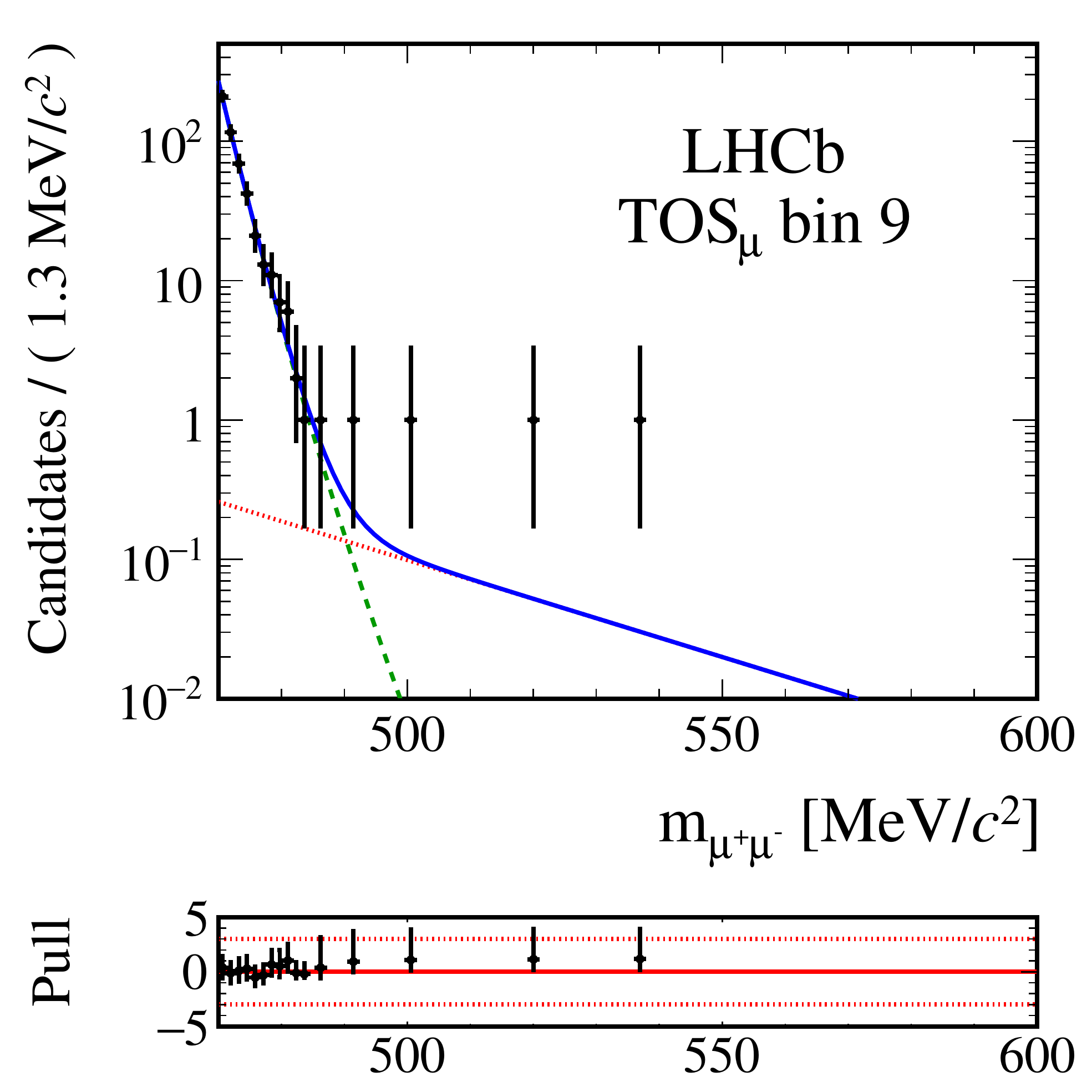}}
        \subfloat{\includegraphics[scale=0.4]{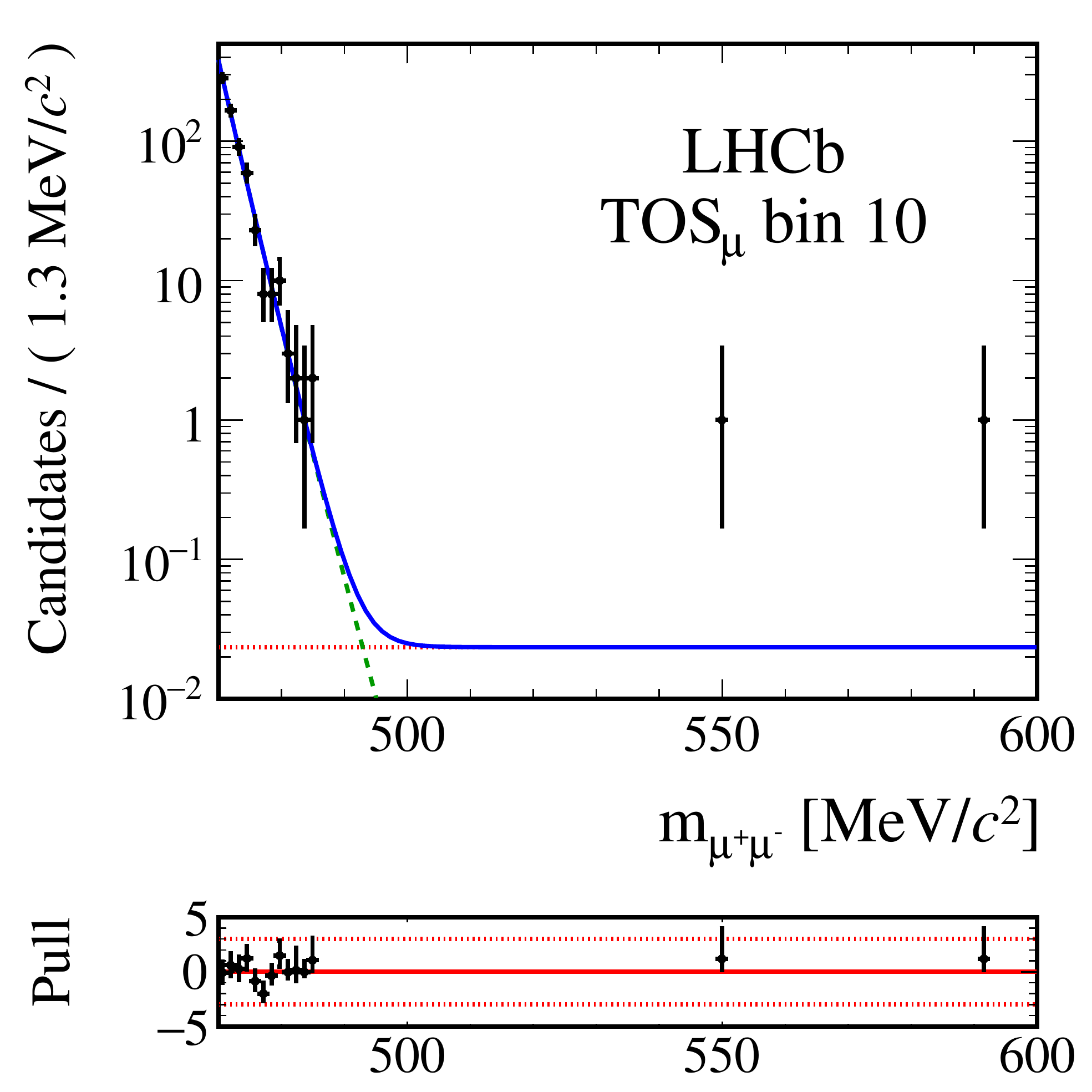}}\\
        \subfloat{\includegraphics[scale=0.4]{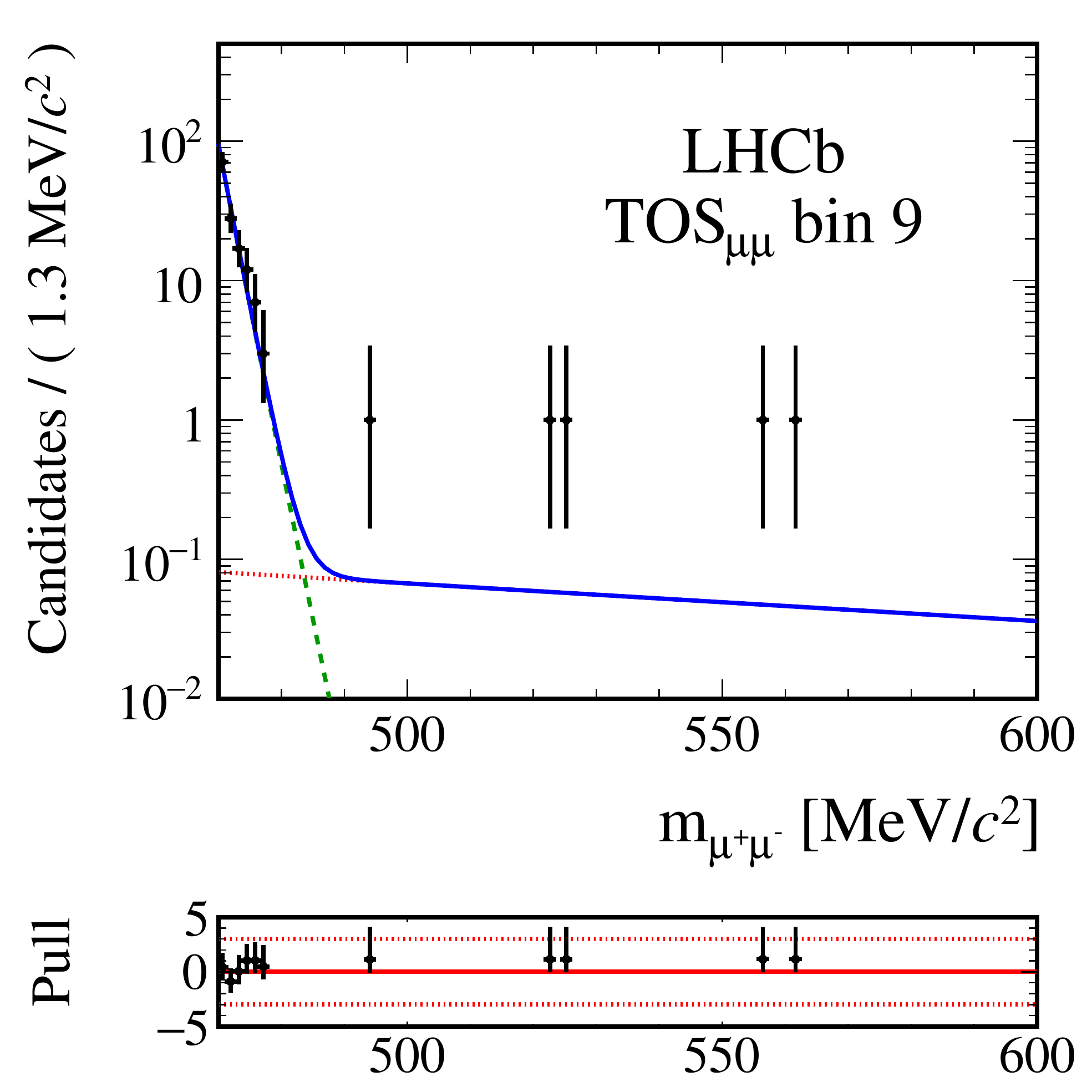}}
        \subfloat{\includegraphics[scale=0.4]{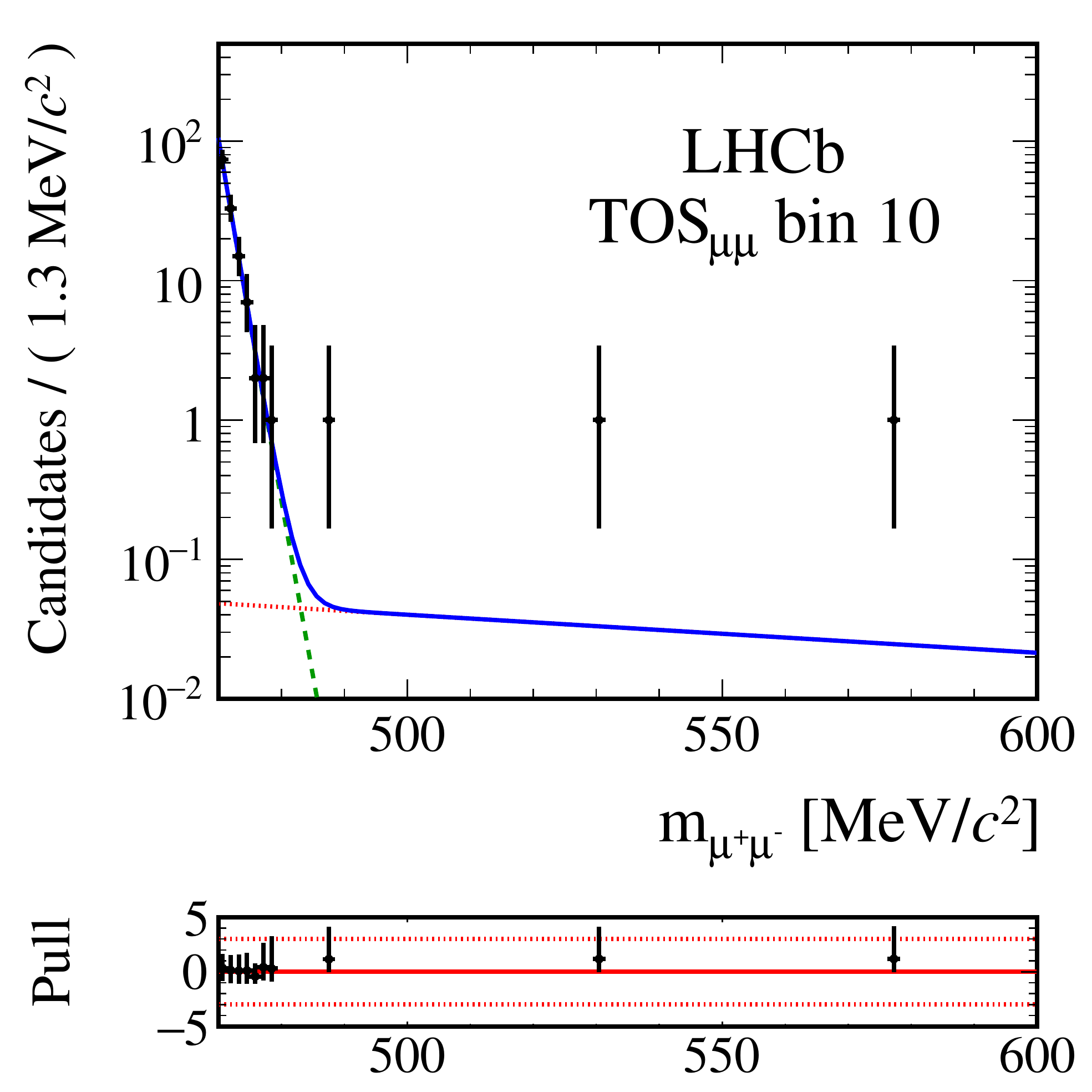}}
        \caption{Fits to the reconstructed kaon mass distributions, for the two most sensitive \BDTcb bins in the two trigger categories, \tosm and \tosmm.
            The fitted model is shown as the solid blue line, while the combinatorial background and \Kspipi double misidentification are overlaid with dotted red and dashed green lines, respectively.
        For each fit, the pulls are shown on the lower smaller plots.\label{fig:fits}}

    \end{center}
\end{figure}

The \mumu mass distribution of the signal candidates is fitted in the range $[470,600]\,\mevcc$ to determine the signal and background yield in each trigger category and \BDTcb bin. 
The model chosen for the signal is a Hypatia distribution~\cite{Hypatia}, the parameters of which are determined from simulation and fixed in the fit to data. 
In the background model, a power law function describes the tail of the double-misidentification background from \Kspipi decays, affecting the mass region below the \Ks mass, while the combinatorial background mass distribution is described by an exponential function.
The background model is validated on simulation, and its parameters are left free in the fit to data to account for possible discrepancies.
An alternative combinatorial background shape, based on a linear function, is used instead of the exponential function to determine a systematic uncertainty due to the choice of the background shape. 
The normalisation channel candidates within the mass region $[460,\,530]\,$\mevcc are counted, leading to $N(\Kspipi)=70\,318\pm 265$.
The \mumu invariant mass distributions for the two highest \BDTcb bins, which exhibit the best signal-to-background ratio and therefore the best sensitivity for a discovery, are shown in Fig.~\ref{fig:fits}.

\begin{figure}[t]
    \begin{center}
      \includegraphics[width=0.6\textwidth]{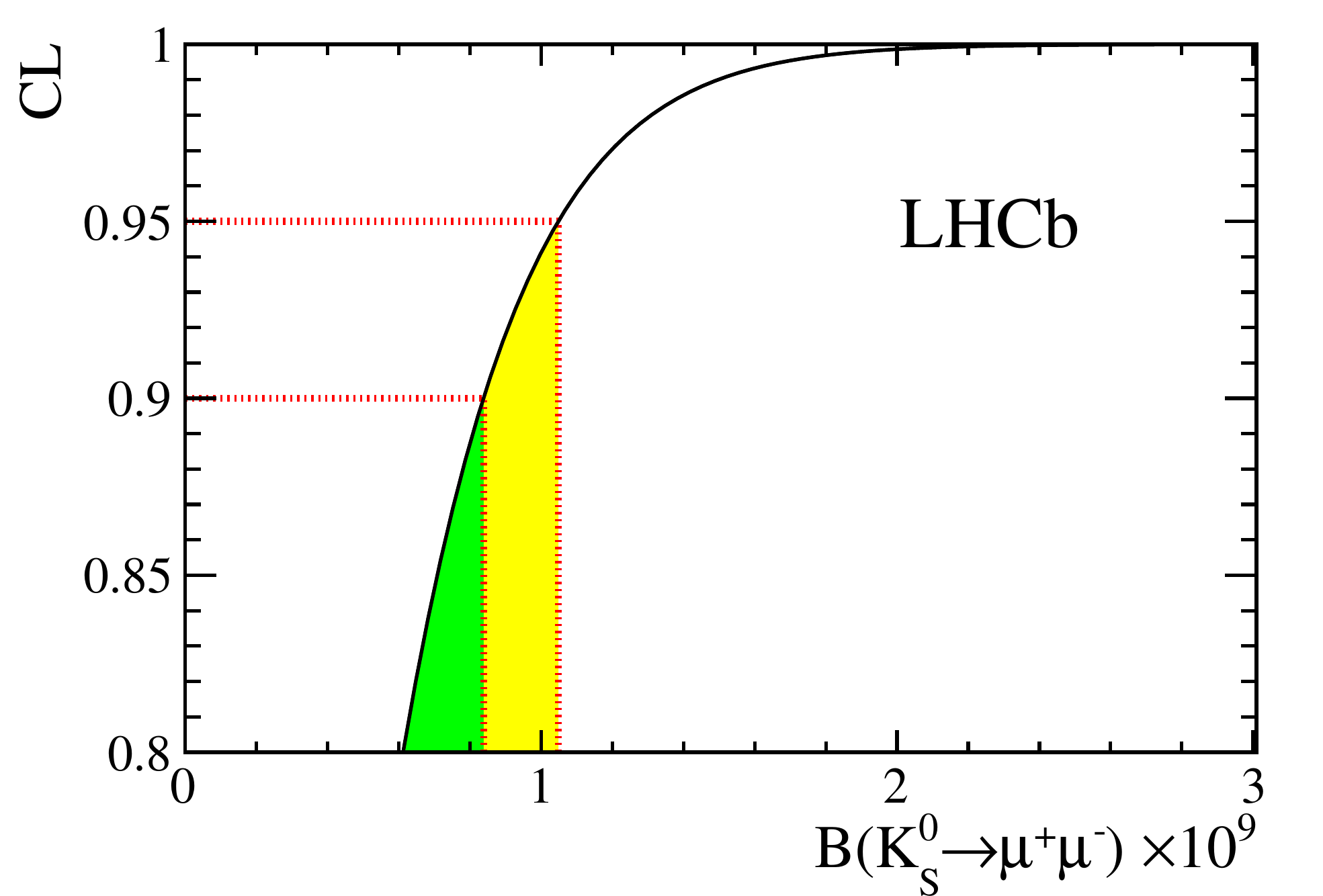}
      \caption{Confidence level of exclusion for each value of the \Ksmumu branching fraction.
          The regions corresponding to $90\%$ and $95\%$ CL are emphasised in green (dark shading) and yellow (light shading), respectively.\label{fig:cllimit}}
    \end{center}
\end{figure}

A simultaneous maximum likelihood fit to the dimuon mass in all \BDTcb bins is performed, using the values of $\alpha_{ij}$ given in Table~\ref{tab:alpha}, to determine the branching fraction. The quoted systematic uncertainties are included  in the likelihood computation as nuisance parameters with Gaussian uncertainties.
A posterior probability is obtained by multiplying the likelihood by a prior density computed from the result based on the 2011 data sample. Limits are obtained by integrating $90\%~(95\%)$ of the area of the posterior probability distribution provided by the fit, as shown in Fig.~\ref{fig:cllimit}. Due to the much larger sensitivity achieved with the 2012 data, the inclusion of the 2011 data result does not have a significant effect on the final limit, and a uniform prior would have provided very similar results.

In conclusion, a search for the \Ksmumu decay based on a data sample corresponding to an integrated luminosity of $3\,\invfb$ of proton-proton collisions, collected by the LHCb experiment at centre-of-mass energies $\sqrt{s}=7$~and~8\tev, allows upper limits to be set on the branching fraction
\begin{equation*}
\BRof\Ksmumu < 0.8~(1.0) \times 10^{-9}~\text{at}~90\%~(95\%)~\text{CL}.
\end{equation*}
This result supersedes the previous upper limit published by \lhcb~\cite{LHCb-PAPER-2012-023}, and represents a factor 11 improvement.

\clearpage

%% file: acknowledgements.tex
\section*{Acknowledgements}

\noindent We express our gratitude to our colleagues in the CERN
accelerator departments for the excellent performance of the LHC. We
thank the technical and administrative staff at the LHCb
institutes. We acknowledge support from CERN and from the national
agencies: CAPES, CNPq, FAPERJ and FINEP (Brazil); MOST and NSFC (China);
CNRS/IN2P3 (France); BMBF, DFG and MPG (Germany); INFN (Italy); 
NWO (The Netherlands); MNiSW and NCN (Poland); MEN/IFA (Romania); 
MinES and FASO (Russia); MinECo (Spain); SNSF and SER (Switzerland); 
NASU (Ukraine); STFC (United Kingdom); NSF (USA).
We acknowledge the computing resources that are provided by CERN, IN2P3 (France), KIT and DESY (Germany), INFN (Italy), SURF (The Netherlands), PIC (Spain), GridPP (United Kingdom), RRCKI and Yandex LLC (Russia), CSCS (Switzerland), IFIN-HH (Romania), CBPF (Brazil), PL-GRID (Poland) and OSC (USA). We are indebted to the communities behind the multiple open 
source software packages on which we depend.
Individual groups or members have received support from AvH Foundation (Germany),
EPLANET, Marie Sk\l{}odowska-Curie Actions and ERC (European Union), 
Conseil G\'{e}n\'{e}ral de Haute-Savoie, Labex ENIGMASS and OCEVU, 
R\'{e}gion Auvergne (France), RFBR and Yandex LLC (Russia), GVA, XuntaGal and GENCAT (Spain), Herchel Smith Fund, The Royal Society, Royal Commission for the Exhibition of 1851 and the Leverhulme Trust (United Kingdom).

%% file: LHCb_Authorship_flat_28-Feb-2017.tex
\centerline{\large\bf LHCb collaboration}
\begin{flushleft}
\small
R.~Aaij$^{40}$,
B.~Adeva$^{39}$,
M.~Adinolfi$^{48}$,
Z.~Ajaltouni$^{5}$,
S.~Akar$^{59}$,
J.~Albrecht$^{10}$,
F.~Alessio$^{40}$,
M.~Alexander$^{53}$,
S.~Ali$^{43}$,
G.~Alkhazov$^{31}$
% ...

{\footnotesize \it
% ...
$ ^{5}$Clermont Universit{\'e}, Universit{\'e} Blaise Pascal, CNRS/IN2P3, LPC, Clermont-Ferrand, France\\
% ...
$ ^{10}$Fakult{\"a}t Physik, Technische Universit{\"a}t Dortmund, Dortmund, Germany\\
% ...
$ ^{31}$Petersburg Nuclear Physics Institute (PNPI), Gatchina, Russia\\
% ...
$ ^{39}$Universidad de Santiago de Compostela, Santiago de Compostela, Spain\\
$ ^{40}$European Organization for Nuclear Research (CERN), Geneva, Switzerland\\
% ...
$ ^{43}$Nikhef National Institute for Subatomic Physics, Amsterdam, The Netherlands\\
% ...
$ ^{48}$H.H. Wills Physics Laboratory, University of Bristol, Bristol, United Kingdom\\
% ...
$ ^{53}$School of Physics and Astronomy, University of Glasgow, Glasgow, United Kingdom\\
% ...
$ ^{59}$University of Cincinnati, Cincinnati, OH, United States\\
% ...
}
\end{flushleft}